\documentclass[aps,prl,twocolumn
]{revtex4-1}
\usepackage{graphicx}
\usepackage{amsmath, amsfonts,amssymb}
\usepackage{color}

\newcommand{\be}{\begin{equation}}
\newcommand{\ee}{\end{equation}}

\newcommand{\bea}{\begin{eqnarray}}
\newcommand{\eea}{\end{eqnarray}}

\newcommand{\Tr}{{\rm Tr}}
\newcommand{\dt}{\partial  _t}

\newcommand{\avg}[1]{{\left\langle #1\right\rangle}}

\newcommand{\Amp}[2]{{\langle #1 | #2 \rangle}}
\newcommand{\Bra}[1]{{| #1 \rangle}}
\newcommand{\Ket}[1]{{| #1 \rangle}}
\newcommand{\doubleAvg}[1]{{\left\langle\! \left\langle#1\right\rangle\!\right\rangle}}

\renewcommand{\vec}[1]{{\bf #1}}

\renewcommand{\epsilon}{\varepsilon}

\newcommand{\addFN}[1]{#1}

\begin{document}
\title{Quantized magnetization density in periodically driven systems}
\author{Frederik Nathan$^1$, Mark S. Rudner$^1$, Netanel H. Lindner$^2$, Erez Berg$^3$  and Gil Refael$^4$}
\affiliation{$^1$Niels Bohr Institute, University of Copenhagen, 2100 Copenhagen, Denmark\\
$^2$Physics Department, Technion, 320003 Haifa, Israel\\
$^3$Department of Condensed Matter Physics, The Weizmann Institute of Science, Rehovot, 76100, Israel\\
$^4$Institute for Quantum Information and Matter, Caltech, Pasadena, California 91125, USA}
\date{\today}
\begin{abstract}
We study micromotion in two-dimensional periodically driven systems in which all bulk Floquet eigenstates are localized by disorder.
We show that this micromotion gives rise to a quantized time-averaged orbital magnetization density in any region completely filled with fermions.
The quantization of magnetization density has a topological origin, and reveals the physical nature of the new phase 
identified in Phys.~Rev.~X {\bf 6}, 021013 (2016).
We thus establish that the 
topological index of this phase can be accessed directly in bulk measurements, 
and propose an experimental protocol to do so using interferometry in cold atom based realizations.
\end{abstract}
\maketitle

Periodic driving was recently introduced as a means for achieving topological phenomena in a wide variety of quantum systems.
Beyond providing new ways to obtain topologically nontrivial band structures~\cite{Niu2007,Oka2009, 
Inoue2010, Lindner2011, 
Lindner2013, Gu11, Floquet_Transport_Kitagawa, Delplace2013, Podolsky2013,  ErhaiPRL, Iadecola2013, Goldman2014, Grushin2014, Kundu2014, Titum2015}, 
periodic driving can give rise to wholly new types of topological phenomena without analogues in equilibrium~\cite{KBRD,WindingNumber, 1D_Majorana, Chiral_1D, 2D_TR, TopologicalSingularities, AFAI, KhemaniPRL2016, RoyHarper2016b, ElseNayakSPT, vonKeyserlingk2016a, Potter2016, RoyHarper2016a,
vonKeyserlingkPRB2016,
vonKeyserlingk2016b,
 NayakFloquetTimeCrystals, TimeCrystalObservation}. 

In a periodically driven system, the unitary Floquet operator acts as a generator of discrete time evolution over each full driving period.
As in non-driven systems, the spectrum and eigenstates of the Floquet operator can be classified according to topology~\cite{KBRD, Oka2009,Lindner2011}. 
However, in addition to the stroboscopic evolution of the system, the {\it micromotion} that takes place within each driving period is crucial for the topological classification of periodically driven systems~\cite{WindingNumber, 1D_Majorana, Chiral_1D, 2D_TR, TopologicalSingularities, RoyHarper2016b, ElseNayakSPT, vonKeyserlingk2016a, Potter2016, RoyHarper2016a}.

Here we uncover a new type of topological quantization phenomenon associated with the micromotion of periodically driven quantum systems.
We focus on periodically driven two-dimensional (2D) lattice systems in which {\it all} bulk Floquet eigenstates are localized by disorder (see Fig.~\ref{fig:Fig1}).
We show that, within a region where all states are occupied, the time-averaged orbital magnetization density $\doubleAvg{{m}}$
  is {\it quantized}: 
 $\doubleAvg{ {m}} = \nu/T$, where $\nu$ is an integer and $T$ is the driving period.
The bulk observable $\doubleAvg{m}$ thus serves as a topological order parameter, characterizing the topologically distinct fully-localized phases found in Ref.~\cite{AFAI}.
We propose a bulk interference measurement to probe this invariant in cold atom systems.


\begin{figure}[t]
\includegraphics[width=0.85\columnwidth]{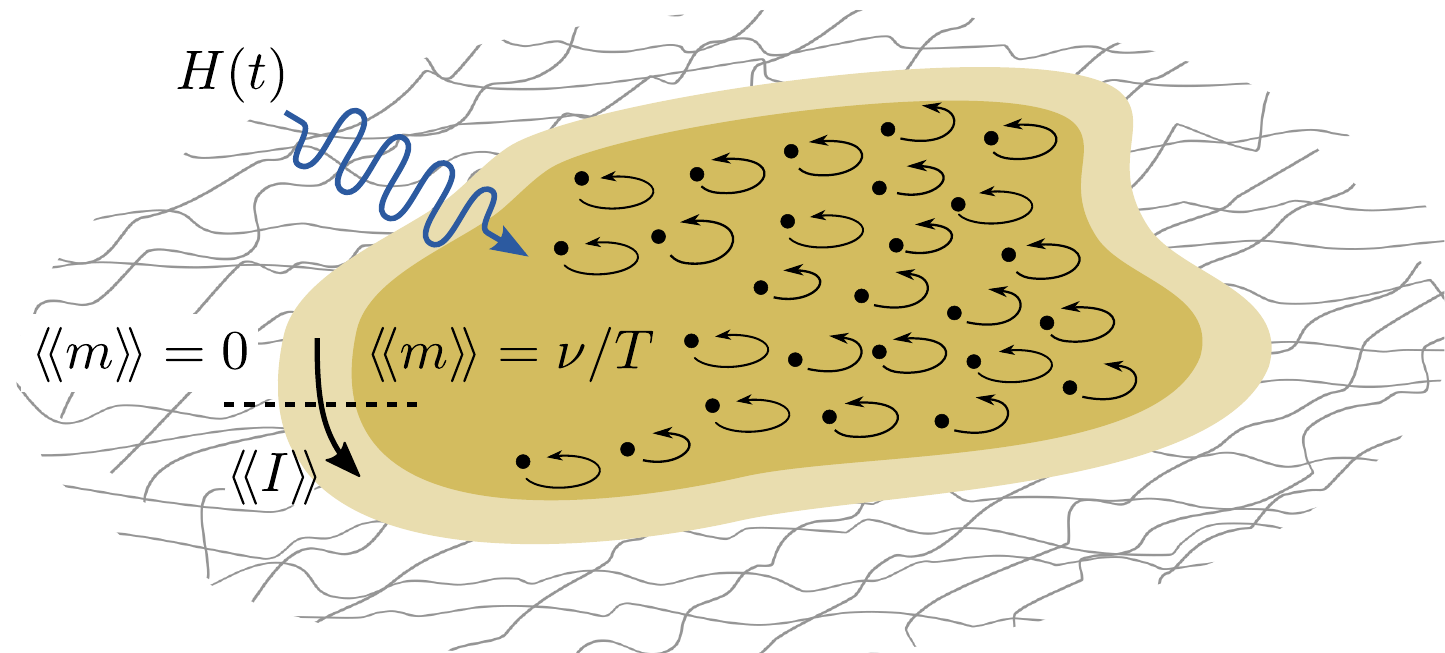}
\caption{Quantized magnetization density in a two-dimensional periodically driven
system where all Floquet eigenstates are localized. 
In a region where all sites are initially occupied (shaded area), the time-averaged orbital magnetization density $\doubleAvg{m}$ is quantized as $\nu/T$, where $\nu$  is an integer and $T$  is the driving period.
A quantized average current $\doubleAvg{I} = \nu/T$  runs along the edge  of the filled region.
}
\vspace{-0.1in}
\label{fig:Fig1}
\end{figure}



Topological invariants are often associated with
quantized response functions. 
Famously, the Hall {\it conductivity} of an insulator is proportional to the 
\addFN{Chern} number~\cite{TKNN}. 
Interestingly, topology in driven systems may directly give rise to quantization of time-averaged observables, such as the pumped {\it current} in the Thouless pump~\cite{ThoulessPump}.
Similarly, the response of magnetization density to changes of chemical potential in a quantum Hall system is quantized when the chemical potential lies in an energy gap~\cite{MacDonaldChapter, Wyder1997, FN:QHEMagnetization}. 
In contrast, here we find 
 quantization of the 
{\it magnetization density} itself. 

For concreteness, we consider a periodically-driven two-dimensional lattice model with one orbital per site.
Dynamics are governed by a time-periodic Hamiltonian $H(t) = H(t + T)$, where $T$ is the driving period.
The periodic driving gives rise to a unitary evolution $U(t) = \mathcal{T} e^{-i\int_0^t\!dt'\, H(t')}$, where $\mathcal{T}$ denotes time ordering.
The spectrum of the Floquet operator $U(T)$, given by $U(T) \Ket{\psi_n(0)} = e^{-i\epsilon_nT}\Ket{\psi_n(0)}$, defines the Floquet eigenstates $\{\Ket{\psi_n(t)}\}$ and their quasienergies $\{\epsilon_n\}$.

We characterize micromotion in this system 
  via the orbital magnetization~\cite{footnote:origin_invariance} 
\be
M(t)= \frac{1}{2}\left(\vec r \times \vec{\dot{r}}(t)\right)\cdot\vec{\hat z},
\label{eq:PR:MagnetizationDefinition}
\ee
where $\vec{\dot r}(t) = -i [\vec r,H(t)]$.
The magnetization operator (\ref{eq:PR:MagnetizationDefinition}) is equivalently expressed as the response of the Hamiltonian to an applied uniform magnetic field ${B}$: $M(t) = -\frac{\partial H(t)}{\partial B}$\cite{SOM}. 
In non-driven systems, the magnetization of a state hence determines the response of its energy to the field: $\Delta E \sim - \vec{M} \cdot \vec{B}$.
In periodically driven systems, a similar relation holds between a Floquet eigenstate's time-averaged magnetization and the response of its quasienergy to an applied magnetic field.
We define $\langle \mathcal{O}\rangle_\tau \equiv \frac{1}{ \tau} \int_0^{\tau} dt\, \langle \psi(t)|\mathcal{O}(t) |\psi(t)\rangle$ as the time-averaged expectation value of an operator $\mathcal{O}(t)$ in the state $\Ket{\psi(t)}$.
The single-period averaged magnetization of a (localized) Floquet eigenstate $\Ket{\psi_n(t)}$ is given by~\cite{SOM, FN:MagneticFieldUnits}: 
\be
\langle M\rangle^{\!(n)}_{T} \equiv \frac{1}{ T} \int_0^T dt\, \langle \psi_n(t)|M(t) |\psi_n(t)\rangle = -\frac{\partial\varepsilon _n}{\partial B}. 
\label{eq:PR:QuasienergyMagneticResponse}
\ee

Using Eqs.~(\ref{eq:PR:MagnetizationDefinition}) and (\ref{eq:PR:QuasienergyMagneticResponse}), we may associate a net magnetization with a single particle in a localized Floquet eigenstate.
It is  useful to define a local time-averaged {\it magnetization density}, associated with each plaquette $p$ of the lattice, that characterizes the response of quasienergy 
to a magnetic flux $\phi_p$ applied locally through plaquette $p$.
We define the magnetization density operator as~\cite{FN:MagnetizationDensityGaugeInvariance}:
\be
m_{p}(t) = -\frac{\partial H(t)}{\partial \phi_p},\quad \phi_p = \int_p d^2 r\, B(\vec{r}),
\label{eq:MagDensityDef1}
\ee
where the integral is taken over the area of plaquette $p$.
The total time-averaged magnetization, $\langle M \rangle _{\tau}$, is  given by the sum of magnetization densities over all plaquettes: $\langle M \rangle_{\tau} = \sum_p  \langle m_{p} \rangle_{\tau} a^2$, where $a$ is the lattice constant. 

The definition of magnetization density in Eq.~(\ref{eq:MagDensityDef1}) applies for both single particle and many-body systems.
In particular,
for a (single or many particle) Floquet  eigenstate $\Ket{\psi(t)}$ with quasienergy $\varepsilon$, 
the time-averaged magnetization density
 is given by $\langle m_{p} \rangle_{T} = -\frac{\partial \varepsilon }{\partial \phi_p}$.

In the continuum, equilibrium magnetization density is related to the current density $\vec j$ through Ampere's law, $\vec j = \nabla \times \vec m$. 
For a (stationary) system on the lattice, 
Ampere's law relates the time-averaged  magnetization densities on adjacent plaquettes $p$  and $q$ to the time-averaged current $\langle I_{pq} \rangle_\tau $ on the bond between them~\cite{SOM}: 
\begin{align}
\langle I_{pq} \rangle_\tau  & = \langle m_{p} \rangle_{\tau}  - \langle m_{q} \rangle_{\tau}.
\label{eq:AmperesLaw}
\end{align}
Here 
 we take positive current  to be counterclockwise with respect to plaquette $p$.

{\it Magnetization in finite droplets.}--- {
We now show that the time-averaged  magnetization density is quantized 
in a finite ``droplet,''  
where all states in a region of linear dimension $R$  are initially occupied while the surrounding region is completely empty (Fig.~\ref{fig:Fig1}).
Specifically, we consider the long-time average of the magnetization density for a plaquette $p$ deep inside the droplet, $\doubleAvg{m_p}$, where $\doubleAvg{\mathcal{O}} \equiv \lim_{\tau \to \infty} \langle \mathcal{O}\rangle_\tau$.
Below we show that $\doubleAvg{m_p}$ 
takes a constant value $\bar m_{\infty}$, up to exponentially small corrections~\cite{footnote-stationary}.
We then show that $\bar{m}_\infty$ is quantized.

Since all Floquet eigenstates are localized, the particle density will only evolve 
significantly in a strip of width $\xi$ around the boundary of the filled region, where $\xi$ is the single-particle localization length of the Floquet eigenstates. 
Hence, the droplet retains its shape up to a smearing of its boundary. At a distance $d \gg \xi$ from this boundary, the density change remains exponentially small in $d/\xi$ at any time.
Within the droplet, all (time-averaged) bond currents therefore vanish: $\langle{I}_{pq}\rangle_\tau = 0$ for all $\tau$.
The magnetization density $\doubleAvg{m_{p}}$ 
 must therefore be the same for all plaquettes deep within the droplet. 

The uniform value of the magnetization density deep within the droplet may depend on the droplet's size.
We note that 
$\doubleAvg{m_p}$ is given by the sum of magnetization contributions from all occupied states that overlap with plaquette $p$.
Therefore, if the droplet size is increased by adding a section of new (filled) sites in a region far away from plaquette $p$, $\doubleAvg{m_p}$ can only change by an exponentially small amount due to the contributions of the tails of the newly added localized states.
Thus, for a plaquette located a distance $d$ from the boundary, we  obtain $\doubleAvg{ m_p} = \bar{m}_\infty + \mathcal{O}(e^{-d/\xi})$, where 
$\bar{m}_\infty$ is the value in the thermodynamic limit.
As we show below, $\bar{m}_\infty$  is quantized. 

Interestingly, a nonzero value of $\bar m_{\infty}$  implies that a  current circulates around the boundary of the droplet. 
The magnetization density drops from the value $\bar m_\infty$  to zero over a distance of order $\xi$  across the droplet's boundary. 
Using Amperes law~\eqref{eq:AmperesLaw}, the total time-averaged current $\doubleAvg{I}$  passing through a cut through this strip (see Fig.~\ref{fig:Fig1}) is $\doubleAvg{I} = \bar{m}_\infty + \mathcal{O}(e^{-R/\xi})$. 

{\it Quantization of magnetization density.}--- 
To prove the quantization of $\bar{m}_\infty$, we consider the  total magnetization $\doubleAvg{M}$  of a droplet of $N$ particles.
On one hand we have $\doubleAvg{M} = \sum'_n \avg{M}^{(n)}_T + \mathcal{O}(N^{1/2})$, where the sum runs over single particle Floquet eigenstates $\Ket{\psi_n}$ with centers localized within the perimeter of the droplet.
The $\mathcal{O}(N^{1/2})$ correction accounts for the partially-occupied Floquet eigenstates near the droplet's boundary.
On the other hand, since the magnetization density deep inside the droplet is constant and given by $\bar{m}_\infty$, we 
 have $\doubleAvg{M} = Na^2 \bar{m}_\infty + \mathcal{O}(N^{1/2})$.
Here $Na^2$ is the total area of the droplet, with the $\mathcal{O}(N^{1/2})$ correction capturing the uncertainty of the area due to its fuzzy boundary.
By equating the expressions for $\doubleAvg{M}$ and taking the $N \rightarrow\infty$ limit, 
we identify 
\be
\label{eq:mavg} \bar{m}_\infty = \lim_{N\rightarrow \infty}\frac{1}{Na^2}\sum^{}_n{\vphantom{\sum}}'\avg{M}^{(n)}_T.
\ee
The quantity $\frac{1}{N}\sum'_n \avg{M}^{(n)}_T$ is simply the average magnetization of  
Floquet eigenstates in the droplet; below, we 
show that this average is quantized in large, fully-localized systems.
To do this, we explicitly compute the average magnetization over all  Floquet eigenstates 
for a fully-localized system on a large torus of area $A = L^2a^2$, where $L^2$ is the number of sites.

For the system on a torus, 
we compute the time-averaged magnetization $\langle M\rangle_T^{(n)}$ of each Floquet eigenstate $\Ket{\psi_n(t)}$ 
 using Eq.~(\ref{eq:PR:QuasienergyMagneticResponse}).
To use the form $\langle M\rangle_T^{(n)} = - \frac{\partial\epsilon_n}{\partial B}$, we must specify how the field $B$ is introduced. 
Crucially, on a torus, the net magnetic flux must be an integer multiple of $\Phi_0$ (the flux quantum)~\cite{footnote:DiracQuantization}; 
consequently, the strength of a {\it uniform} field cannot be varied continuously.
However, for $\xi/L \ll 1$, 
we may use $\langle M\rangle_T^{(n)} = -\frac{\partial \epsilon_n}{\partial B} + \mathcal{O}(e^{-L/\xi})$, 
where $\epsilon_n(B)$ is the quasienergy of state $\Ket{\psi_n}$ in the presence of a {\it locally} uniform 
 magnetic field, 
of strength $B$ 
 within the support region of $\Ket{\psi_n}$, 
but zero net flux through the torus.
The  $\mathcal{O}(e^{-L/\xi})$ correction arises from the non-uniformity of the field, which is concentrated where the wave function is exponentially small. 

To evaluate the average magnetization of localized Floquet eigenstates, 
$\frac{1}{L^2}\sum_n \avg{M}_T^{(n)} =-\frac{1}{L^2}\sum_n \frac{\partial \epsilon_n}{\partial B} $, we examine the Floquet operator $U(T)$ in the presence of a global uniform magnetic field of strength 
 $B_0= 2\pi/A$, 
corresponding to precisely one flux quantum piercing the torus.
For large $A$, the quasienergy in the uniform field $B_0$ is equal to that in the locally  uniform field described above (with $B = B_0$), 
up to an exponentially small correction in $L/\xi$.
Moreover, for small field strengths, $\frac{\partial\epsilon_n}{\partial B}$ is well approximated by a finite difference, such that~\cite{footnote:labeling}: 
\be
\langle M \rangle^{\!(n)}_{T} = -[\varepsilon_n (B_0) - \varepsilon_n (0)]/B_0  + \mathcal O (1/A).
\label{eq:QuasienergyMagnetizationExpansion}
\ee
The $\mathcal{O}(1/A)$ correction accounts for the error in discretizing the derivative. 

Using Eq.~(\ref{eq:QuasienergyMagnetizationExpansion}), we can access $\sum_n \avg{M}_T^{(n)}$ directly via the determinant of the system's Floquet operator~\cite{TopologicalSingularities}, $|U(T)|$.
Writing $\log |U(T)| = \int_0^T dt\, \partial _t \log|U(t)|$, we use the identity $\partial _t \log|U(t)|=\Tr\left[ U^\dagger(t)\dt U(t)\right]$, together with $\partial _t U(t) = -i H(t) U(t)$,  and find~\cite{footnote:InitialCondition}
\be
\label{eq:LogDetU}\log |U(T)| = - i \int_0^T dt\,  \Tr \left[H(t) \right].
\ee
When a magnetic field is introduced, the hopping amplitudes between sites of the lattice acquire additional Peierl's phases:
$H_{ab} \rightarrow H_{ab}e^{i\theta_{ab}}$. 
In the position basis, the magnetic field thus only affects the {\it off-diagonal} elements of the Hamiltonian, and we conclude that ${\rm Tr}[H(t)]$ and hence $|U(T)|$ are independent of magnetic field.
Using $|U(T)|=e^{-i\sum_n \varepsilon_n T}$, we  find
\be
\sum_n \varepsilon_n(B_0)  = \sum_n \varepsilon_n(0) - \frac{2\pi \nu}{T},
\label{eq:PR:QuasienergyRelation}
\ee
where $\nu$  is an integer. 

Recall that $\bar m_\infty$, the magnetization density in a filled droplet is obtained from the average magnetization of the Floquet eigenstates in the droplet, see Eq.~(\ref{eq:mavg}).
The torus geometry discussed above allows us to compute this average in the thermodynamic limit. 
Using Eqs.~(\ref{eq:QuasienergyMagnetizationExpansion}) and (\ref{eq:PR:QuasienergyRelation}) we obtain $\frac{1}{L^2} \sum_n \langle M \rangle^{(n)}_T=  \frac{2\pi\nu}{L^2B_0 T}$~\cite{footnote:notorusmag}.
Comparing to Eq.~(\ref{eq:mavg}), we find:
\be
\bar{m}_\infty = \frac{\nu}{T}.
\label{eq: minfinity}
\ee
Remarkably, this quantization has a topological origin.
As we show in the SOM~\cite{SOM}, the integer $\nu$ is equal to the winding number invariant characterizing the Anomalous Floquet-Anderson Insulator (AFAI) phase, introduced in Ref.~\cite{AFAI}.
The 
 magnetization density thus 
serves as a bulk topological order parameter that characterizes 
 distinct fully-localized Floquet phases.
Note that the emergence of a non-zero, quantized 
 magnetization density 
 is a unique dynamical phenomenon, with no counterpart in non-driven systems:
for static systems, Eq.~\eqref{eq: minfinity} must hold for all values of $T$, which requires $\nu=0$~\cite{footnote:Ref11Comments}.


\begin{figure}[t]
\includegraphics[width=0.85\columnwidth]{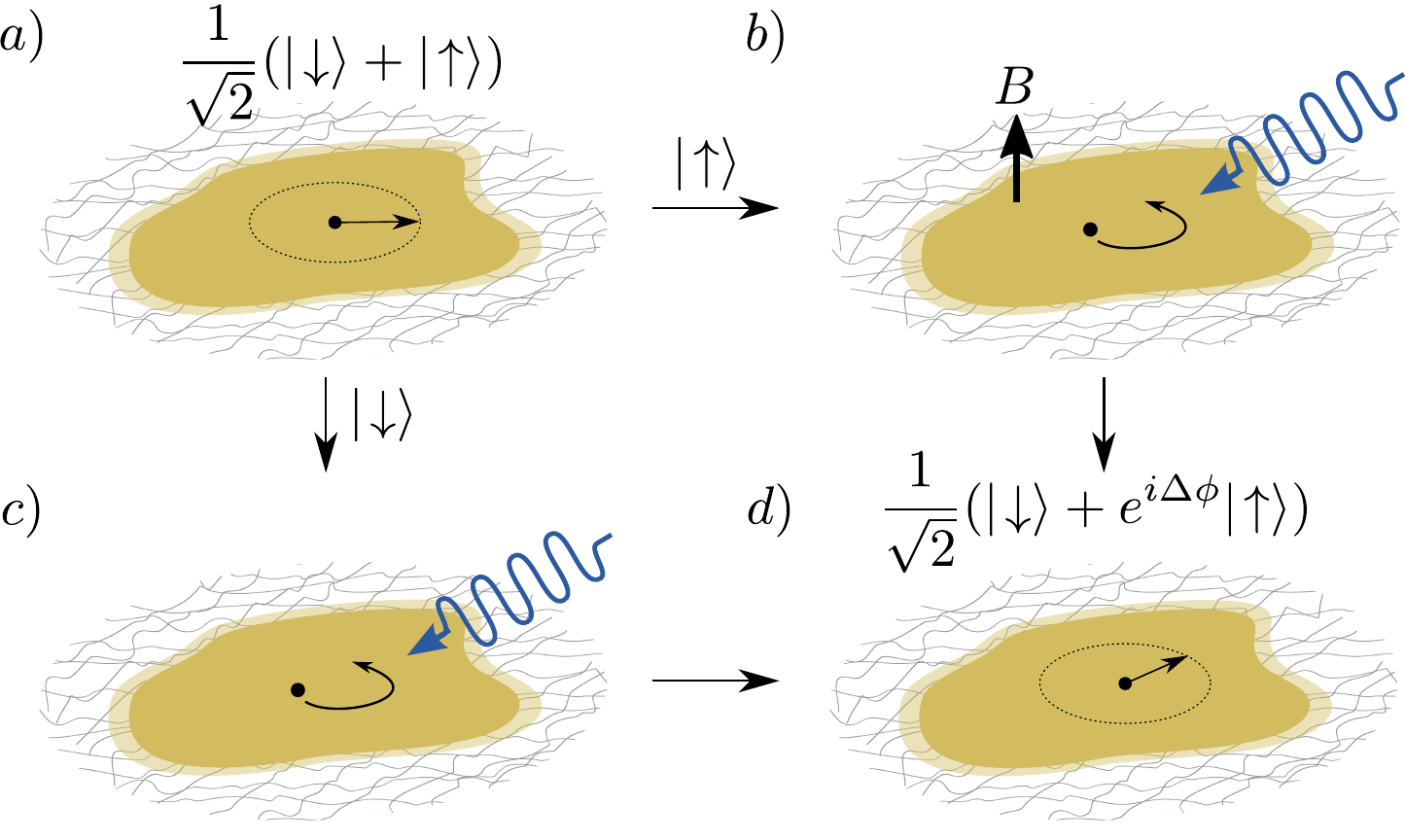} 
\caption{Interferometric measurement of quantized orbital magnetization density in a cold-atom system.
a) The system is 
 prepared by filling 
 a region of an optical lattice with spin-1/2 atoms 
fully polarized along $x$.
The system is evolved with a spin-independent periodic driving Hamiltonian, plus a weak spin-dependent uniform synthetic magnetic field
\addFN{ that only affects the $|\!\!\uparrow\rangle$ component of the system's wave function \addFN{(b)}, while the  $|\!\!\downarrow\rangle$  component is unaffected \addFN{(c)}. 
}
d) The spin-dependent field gives rise to a 
phase-difference $\Delta \phi$  between the $|\!\!\uparrow\rangle$ and $|\!\!\downarrow\rangle$  components of each  atom's wave function.
The phase shift yields a net $y$-polarization of total spin, proportional to the system's time-averaged magnetization.
}
\vspace{-0.2in}
\label{fig:ColdAtomExperiment}
\end{figure}
{\it Interferometric probe of quantized magnetization.}--- 
We now outline an interferometric scheme for measuring the spatially averaged magnetization density  $\doubleAvg{\overline m} = \doubleAvg{M}/A_{\rm filled}$ of a cloud of fermionic cold atoms in an optical lattice (see Fig.~\ref{fig:ColdAtomExperiment}), where $A_{\rm filled}$ is the area of the initially filled region. 
We thus offer a direct probe to measure the bulk topological invariant of the AFAI.

Consider an atom traversing a closed trajectory in the presence of a weak magnetic field $B$.
Semiclassically, the wave-function picks up an additional phase shift $\Delta \phi = BA_{\rm orb}$ due to the field, where $A_{\rm orb}$ is the area enclosed by the orbit~\cite{FN:WaveFunctionOverlap}.
Correspondingly, a simple quantum mechanical calculation~\cite{SOM} shows that
the phase shift acquired by an atom in Floquet eigenstate $\Ket{\psi_n(t)}$ over a full driving period is proportional to the state's magnetization, 
$
\Delta \phi_n = \langle M \rangle^{\!(n)}_T  BT.
$
Using this phase shift, the magnetization of a cloud of atoms can be measured in a Ramsey-type interference experiment in a situation where the atoms have two internal (``spin'') states $|\!\!\uparrow\rangle$  and $|\!\!\downarrow\rangle$. 
First, 
the system should be 
prepared
 by completely filling a region of known area, $A_{\rm filled}$, with atoms fully spin-polarized along the ``$x$''-direction, $|\psi(0)\rangle  \propto (|\!\!\uparrow\rangle + |\!\!\downarrow\rangle)/\sqrt{2}$, 
(Fig.~\ref{fig:ColdAtomExperiment}a). 
The system should then be evolved with the driving Hamiltonian 
to allow the particle density to reach a steady profile~\cite{footnote:Initialization}, as in Fig.~\ref{fig:Numerics}a.
To perform the measurement, the cloud of atoms is then evolved through $N$ driving  periods 
 in the presence of a weak spin-dependent orbital effective magnetic field $B$ (Figs.~\ref{fig:ColdAtomExperiment}bc), which, e.g., acts only on the $|\!\!\uparrow\rangle$ species.
Through the evolution, the $|\!\!\uparrow\rangle$ component of each atom's wave function  gains a phase shift  relative to the $|\!\!\downarrow\rangle$ component,
yielding a nonzero average 
$y$-spin per particle, $\langle \overline{\sigma}_y\rangle$, 
(Fig.~\ref{fig:ColdAtomExperiment}d).
For small precession angles, the average $y$-spin 
after $N$ periods is given by $\avg{\overline{\sigma}_y(NT)} \equiv \Omega_{NT} Ba^2 NT$, with~\cite{SOM} 
\be
\Omega_{NT} =   \doubleAvg{\overline m} +\frac{1}{NT} \mathcal{O} \left(\frac{\xi^{3/2}}{aR^{1/2}}\right) + \mathcal O (B).
\label{eq:AverageMagnetizationDensity}
\ee

 Importantly, the second term vanishes in the long time limit (and  scales to zero at finite times for large systems), thus revealing the quantized magnetization density~\cite{SOM}.


\begin{figure}
\includegraphics[width=0.85\columnwidth]{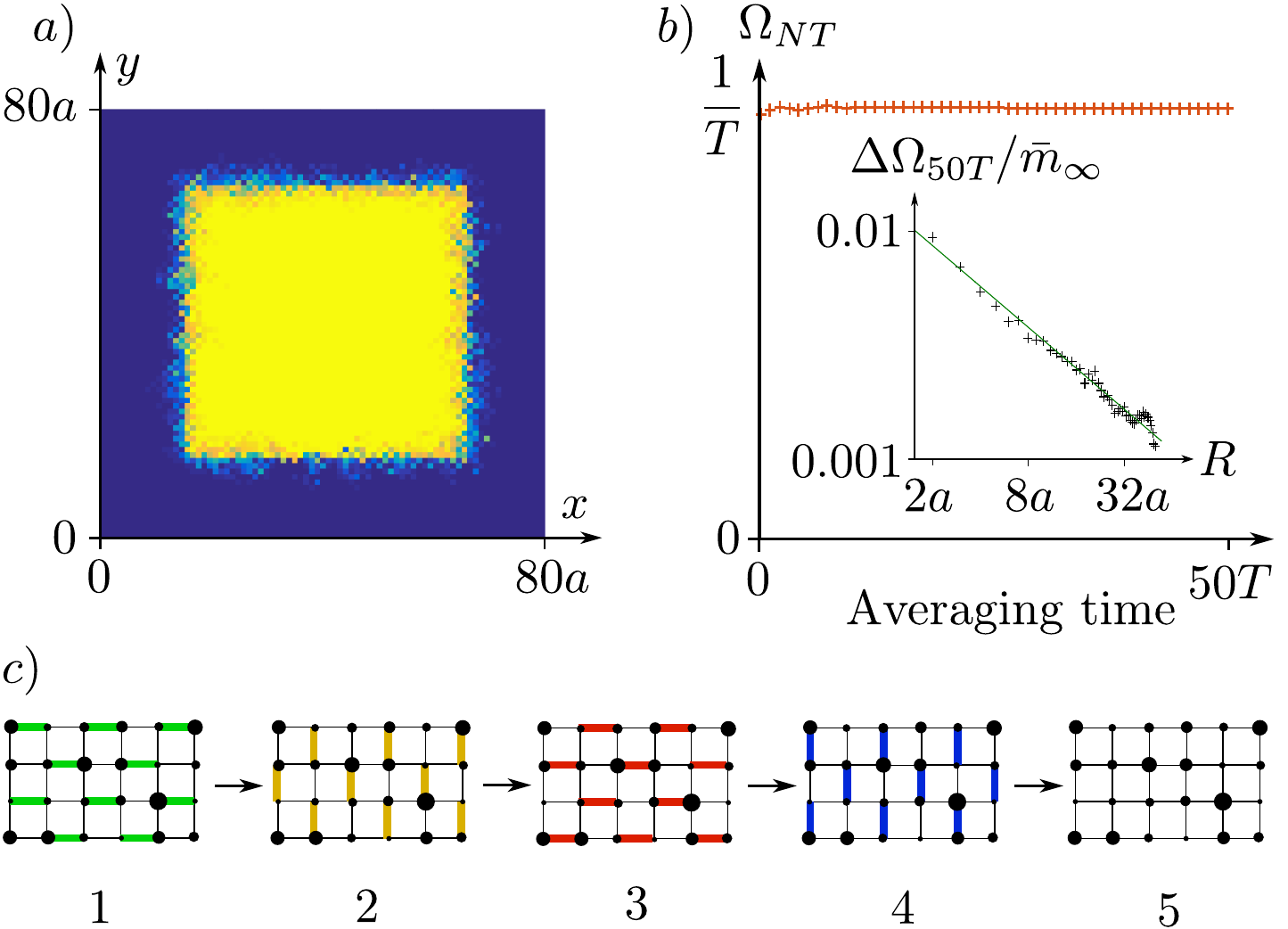}
\caption{
a)
Particle density in the system after $20$ driving periods, for an initially filled $50\times50$ square of sites.
b) 
Normalized growth rate $\Omega_{NT}$ 
of the average $y$-spin per atom [see text above Eq.~(\ref{eq:AverageMagnetizationDensity})]. 
The long-time-averaged magnetization density $\doubleAvg{\overline m}$  is extracted from the saturation value at long times.
{\it Inset}: Deviation $\Delta\Omega_{50T}$ 
of $\Omega_{50T}$ 
from the quantized value $\bar{m}_\infty$ 
{\it vs.}~droplet size $R$.  
The value of $\Delta \Omega_{50T}$ is obtained as \addFN{a} RMS average of $ \Omega_{50T} - \bar{m}_\infty$ over $100$ disorder realizations.
c)
Depiction of the tight-binding model.}
\label{fig:Numerics}
\end{figure}

{\it Numerical results.}---
We simulated the experimental protocol outlined above
using a tight-binding model on a  two-dimensional bipartite square lattice, with Hamiltonian 
$H(t)=H_{\rm clean}(t) + V_{\rm disorder}$.
The Hamiltonian $H_{\rm clean}(t)$ was considered in Ref.~\cite{WindingNumber}, and is of the form 
\be
H_{\rm clean}(t) = \sum_{ \vec r \in A} \sum_{n=1}^4 J_n(t) (c^\dagger_{\vec r + \vec b_n}c_{\vec r} + h.c.), 
\ee
where $c_\vec r$  is the fermionic annihilation operator on the lattice site with coordinate $\vec r$, and the first sum runs over sites $\vec r$  on sublattice $A$.
The vectors $\{\vec b_n\}$  are given by $\vec b_1=-\vec b_3= (a,0)$  and $\vec b_2 =- \vec b_4 = (0,a)$, where $a$  is the lattice constant.
The  driving period is divided into five segments of equal length $T/5$.
In the  $n$th segment ($n\le 4$), $J_{n}(t)= J$, while all other hopping amplitudes are set to zero; in the $5$th segment all hopping amplitudes are set to zero (see Fig.~\ref{fig:Numerics}c).
We introduce disorder 
through a time-independent 
potential $V_{\rm disorder}= \sum_{\vec r} w_{\vec r} c^\dagger_{\vec r}c_{\vec r}$,
where the sum  runs over all sites
, and
the  on-site energies $\{w_{\vec r}\}$ are randomly drawn from a uniform distribution in the interval $[-W,W]$.
The model has hopping amplitude $J$  and disorder strength $W$  both set to $2.5\pi /  T $.
This brings  the system well into the AFAI phase, for which we expect $\bar{m}_\infty = 1/T$~\cite{SOM}. 

To find the magnetization density of the system, we
consider a single disorder realization on a lattice of $80\times 80$  sites and open  boundary conditions.
We initially 
fill  a region of $50\times 50$ sites (i.e., $R = 50$)   centered in the middle of the lattice, and prepare the state by evolving it for $20$ driving periods at zero magnetic field (see Fig.~\ref{fig:Numerics}a).
For further times ranging from $0$  to $50T$ we evolve the system in the presence of a spin-dependent magnetic field of strength $Ba^2= 2\pi \cdot 10^{-4}$. 
We extract the 
spatially averaged  magnetization density $\doubleAvg{\overline{m}}$ from the long-time limit of the normalized growth rate $\Omega_{NT}$ of 
 average $y$-spin per atom, $\avg{\overline{\sigma}_y(NT)}$. 
$\Omega_{NT}$ 
rapidly converges (up to a finite-size correction) 
to the quantized value of the magnetization density, $1/T$,  
reaching 
$0.9998$ after $100$  periods (see Fig.~\ref{fig:Numerics}b and SOM). 
The inset in Fig.~\ref{fig:Numerics}b shows  the deviation of $\Omega_{50T}$  from the quantized value $\bar{m}_\infty = 1/T$ for various sizes of the droplet, taken as  a root-mean-square average over $100$  disorder realizations at each system size. 
We find a power law decay of the fluctuations with system size, $\Delta \Omega_{50T} \sim R^{-0.55}$.

{\it Discussion.}---
Here we showed that the orbital magnetization density is quantized in fully-filled regions of localized Floquet systems. 
We then proposed an experimental scheme for measuring the quantized magnetization density in cold atomic systems.

We  derived the quantization of magnetization density within a tight-binding model with one ($s$-type) orbital per site.
This means that each on-site orbital does not carry any intrinsic magnetization.
In the continuum, small non-quantized contributions to the magnetization density may arise due to mixing with higher bands.
Such contributions are strongly suppressed when the driving is adiabatic with respect to the gap to higher bands, and the lattice is very deep such that the gap is large compared to the bandwidth~\cite{SOM}.


It is natural to expect that our results will hold also in the presence of interactions, given that the system is strongly disordered and hence may be many-body localized.
Recently,  progress has been made in constructing interacting analogues of the AFAI~\cite{AshvinInteracting, RoyAFAI}. The fate of the magnetization in the presence of interactions remains an open direction of investigation. 

\begin{acknowledgements}
{\it Acknowledgements.---} M.R. gratefully acknowledges the Villum Foundation and the People Programme (Marie Curie Actions) of the European Unions Seventh Framework Programme (FP7/2007-2013) under \addFN{Research Executive Agency (REA)}  
 grant agreement PIIF-GA-2013-627838 for support.
N.L. and E.B. acknowledge financial support from the European Research Council (ERC) under the European Unions Horizon 2020 research and innovation programme (grant agreement No 639172).
\addFN{N. L. acknowledges support from the People Programme (Marie Curie Actions) of the European Union's Seventh Framework Programme (No. FP7/2007-2013) under REA Grant Agreement No. 631696, and from the Israeli Center of Research Excellence (I-CORE) ``Circle of Light''.}
G. R. is grateful for support from the \addFN{National Science Foundation (NSF)} through Grant No. DMR-1410435, the Institute of Quantum Information and Matter, an \addFN{National Science Foundation} 
 Frontier center funded by the Gordon and Betty Moore Foundation, and the Packard Foundation, and further thanks the Aspen Center for Physics for their hospitality.
\end{acknowledgements}

\onecolumngrid

\newpage 
\newpage 
\author{Frederik Nathan$^1$, Mark S. Rudner$^1$, Netanel H. Lindner$^2$, Erez Berg$^3$  and Gil Refael$^4$}
\affiliation{$^1$Niels Bohr Institute, University of Copenhagen, 2100 Copenhagen, Denmark\\
$^2$Physics Department, Technion, 320003 Haifa, Israel\\
$^3$Department of Condensed Matter Physics, The Weizmann Institute of Science, Rehovot, 76100, Israel\\
$^4$Institute for Quantum Information and Matter, Caltech, Pasadena, California 91125, USA}
\begin{center}
\textbf{\large Supplementary Material: \\
Quantized magnetization density in periodically driven systems} 
\vspace{0.2cm}
\end{center}
\maketitle
\twocolumngrid

\section{Magnetization as the response of quasienergy to a magnetic field}
\label{sec:MagnetizationHamiltonianRelationship}
Here we derive Eq.~(2) in the main text, showing that the single-period averaged magnetization  $\langle M\rangle^{\!(n)}_T$ of a Floquet state $|\psi_n\rangle$ with quasienergy $\varepsilon_n$ is given by the response of its quasienergy to an applied ``probing'' uniform magnetic field, $B$: $\langle M\rangle^{\!(n)}_T = -\frac{\partial \varepsilon_n}{\partial B}$.
(Note that, in addition to the probing field $B$, a nontrivial field $B_0(\vec{r},t)$ may already be present in the system.)
Throughout this work the magnetic field is given in units of [1/Area], such that the flux quantum has value $2\pi$.  

As a first step, we note that $\frac{\partial \varepsilon_n}{\partial B}$  can be written as 
\begin{align} 
\frac{\partial \varepsilon_n}{\partial B} = \frac{i}{T}\langle \psi_n|\left(U^\dagger(T)\frac{\partial}{\partial B}U(T)\right)|\psi_n\rangle.
\label{eq:QEDeriv}
\end{align}
This relation can be checked using the spectral decomposition $U(T)=\sum_n|\psi_n\rangle\langle\psi_n|e^{-i\varepsilon_n T}$, together with the identity $\langle \psi_n|\frac{\partial }{\partial B} |\psi_n\rangle + \frac{\partial  }{\partial B}\big[\langle \psi_n |\big]| \psi_n\rangle = 0$.
Here $\frac{\partial }{\partial B}|\psi_n\rangle$  measures the change of Floquet state $\Ket{\psi_n}$ when a uniform magnetic field $B$ is introduced to the system.

We now use $U(T)=\mathcal T e^{-i\int_0^T dt H(t)}$ to obtain 
\begin{align} 
U^\dagger(T)\frac{\partial}{\partial B}U(T)
=-i \int_0^T\!\! dt\, U^\dagger(t) \frac{\partial H(t)}{\partial B}U(t) .
\label{eq:QEDeriv2}
\end{align}
Hence, substituting back into Eq.~(\ref{eq:QEDeriv}), we get
\be 
\frac{\partial \varepsilon_n}{\partial B} 
= \frac{1}{T} \int_0^T
\langle \psi_n(t)|
\frac{\partial H(t)}{\partial B}|\psi_n(t)\rangle, 
\label{eq:QEHamiltonianResponse}
\ee
where $|\psi_n(t)\rangle = U(t)|\psi_n\rangle$ is the time-evolved Floquet eigenstate at time $t$. 

%
%

What is the nature of the operator $\frac{\partial H}{\partial B}$?
By analogy to equilibrium systems, clearly it is suggestive of magnetization.
However, similar to the magnetization density operator $m_p$ discussed in the main text, the operator $\frac{\partial H}{\partial B}$ is gauge-dependent.
Nonetheless, expectation values of $\frac{\partial H}{\partial B}$ taken in {\it stationary states} are in fact gauge invariant, and therefore physical (see next section).
The stationarity condition is satisfied for the full-period average of $\frac{\partial H}{\partial B}$ in a Floquet state, as appears on the right hand side of Eq.~\eqref{eq:QEHamiltonianResponse}.
Indeed this must be the case, since the quantity $\frac{\partial\varepsilon_n}{\partial B}$ on the left hand side is itself gauge-invariant.

To obtain an expression for $\frac{\partial H(t)}{\partial B}$, we consider the change of the Hamiltonian when the small uniform probing magnetic field $B$ is introduced. 
In this case,  the matrix elements $H_{ab}(t)$  of the  Hamiltonian in the lattice site basis (here $a,b$  refer to lattice site indices)  acquire Peierl's phases: $H_{ab}(t)\rightarrow H_{ab}(t) e^{i\int_{\vec r_b}^{\vec r_a} d\vec r \cdot \vec A(\vec r)}$, where the contour of integration is a straight line from site $b$  to site $a$ and $\vec{B} = \nabla \times \vec{A}$.
Given that the result of Eq.~\eqref{eq:QEHamiltonianResponse} is gauge-independent,  we work in the symmetric gauge below.
This gauge choice highlights the direct relation to the magnetization defined in Eq.~(1) of the main text. 
In the symmetric gauge, a uniform perpendicular ``probing'' magnetic field $B$ is produced by the vector potential $\vec A(\vec r) = \frac{B}{2} \vec{\hat z} \times \vec r$.
Using the identity $A\cdot (B\times C)= B\cdot (C \times A)$,  we thus obtain the following modification of $H_{ab}(t)$  due to the probe field $B$ : 
\begin{align*}
H_{ab}(t) & \rightarrow H_{ab}(t)\exp\left[\frac{i B}{2} \int_{\vec r_b}^{\vec r_a} d\vec r \cdot (\hat{\vec z} \times \vec r) \right] \\
&= H_{ab}(t)\exp\left[\frac{i B}{2}\hat{\vec z} \cdot \left(\int_{\vec r_b}^{\vec r_a}\vec r \times d{\vec r} \right) \right]\\
&= H_{ab}(t)\exp\left[\frac{i B}{2}\hat {\vec z} \cdot \left( \vec r_a \times (\vec r_a - \vec r_b)\right)\right].
\end{align*}
Here we used that $\vec r_a \times (\vec r_a - \vec r_b) = \vec r_b \times (\vec r_a - \vec r_b)$.

Taking the derivative of $H_{ab}(t)$  with respect to the probe field strength $B$, we obtain
\begin{align}
\frac{\partial H_{ab}(t)}{\partial B} &=  \frac{i}{2}H_{ab}(t)  \left(\vec r_a \times (\vec r_a - \vec r_b)\right)\cdot \vec{\hat z}. 
\label{eq:pre_commutatorForm}
\end{align}
This structure of the matrix elements of $H$ implies that 
\be 
\frac{\partial H(t) }{\partial B} = \frac{i}{2} \left(\vec r\times [\vec r, H(t)]\right)\cdot \vec{\hat z}.
\label{eq:commutatorForm}
\ee
Equation (\ref{eq:commutatorForm}) can be verified by taking a matrix element with $\Bra{a}$ and $\Ket{b}$ on the left and right, respectively, and comparing with Eq.~(\ref{eq:pre_commutatorForm}).
Comparing with Eq.~(1) of the main text, and using $\dot{\vec{r}}(t) = -i[\vec{r},H]$, we identify the right hand side above as minus the magnetization, $-M(t)$.
Substituting this result into Eq.~\eqref{eq:QEHamiltonianResponse}, we obtain Eq.~(2) in the main text. 

\section{Gauge invariance of magnetization density} 
\label{sec:GaugeInvariance}
Here we show that the 
magnetization density operator $ m_p (t)$, defined in Eq.~(3) of the main text, yields gauge-independent time-averaged expectation values if and only if the density is stationary over the averaging interval $\tau$, i.e., $\langle \dot{\rho}\rangle_\tau =0$.
In this case, we furthermore show that the magnetization density obeys the lattice version of Ampere's law given in Eq.~(4) of the main text.

%

In the presence of a magnetic flux $\phi_p$ piercing through plaquette $p$, the matrix elements of the Hamiltonian in the lattice site basis are given by $H_{ab}(\phi_p)=e^{iA_{ab}(\phi_p) }H_{ab}(\phi_p = 0)$. (Here we work in units where the lattice constant is $1$). 
Here the vector potential $\{A_{ab}(\phi_p)\}$  should have the following property: 
for a sequence of sites $(a_1,a_2,\ldots a_N)$  forming a closed counterclockwise loop on the lattice, the phase $\sum_{n=1}^N A_{a_{n+1}a_n}(\phi_p)$ should equal $\phi_p$ if the loop encloses the plaquette $p$, while the sum should vanish otherwise (here we set $a_{N+1} = a_1$). 
The magnetization density operator is then given by
\be 
m_p(t) =-\frac{\partial H(t)}{\partial \phi_p} =  - \sum_{\langle a,b\rangle}\frac{\partial H(t)}{\partial A_{ab}}\frac{\partial A_{ab}}{\partial \phi_p},
\label{eq:mpDeriv}
\ee
where the sum runs over all pairs of sites on the lattice connected by bonds. 

We note that there is a gauge freedom in choosing $A_{ab}(\phi_p)$:  if the vector potential $\{A_{ab}(\phi_p)\}$  results in a flux $\phi_p$  on plaquette $p$, then so will a vector potential $\{A'_{ab}(\phi_p)\}$ that satisfies
\be 
A'_{ab}(\phi_p) =A_{ab}(\phi_p)+f_a(\phi_p)-f_b(\phi_p),
\ee
where  $\{f_a(\phi)\}$  can be any set of scalar functions.

In order for $\langle m_p\rangle_\tau$  to be gauge-invariant, the time-averaged expectation value of the right hand side of Eq.~(\ref{eq:mpDeriv}) should remain unchanged if we replace $A_{ab}$  with $A_{ab}'$. 
%
%
In order for this to be satisfied, we must have
\be 
\sum_{\langle a,b\rangle}\left\langle \frac{\partial H}{\partial A_{ab}}\right\rangle_\tau(g_a-g_b)=0,
\label{eq:gaugeinvariance}
\ee
where  $\{g_a =\frac{\partial f_a}{\partial \phi}\big|_{\phi=0}\}$  are arbitrary coefficients.
Equation (\ref{eq:gaugeinvariance}) is satisfied if we require that the net current flowing into or out of every site $a$ on the lattice vanishes: 
\be 
\sum_{b\in {\rm n.n.}(a)}\!\!\left\langle I_{ab} \right\rangle_\tau=0,\quad I_{ab}(t) = -\frac{\partial H(t)}{\partial A_{ab}}.
\label{eq:divI}
\ee
Here the sum runs over all sites $b$  that are connected with a bond to site $a$.
It is trivial to see that this condition ensures that the sum over terms proportional to $g_a$ in Eq.~(\ref{eq:gaugeinvariance}) vanishes.
The vanishing of the sum over terms proportional to $g_b$ follows by relabeling.

The sum on the left hand side of Eq.~(\ref{eq:divI}) gives the net current flowing into site $a$, which is equal to the rate of change of density: $\sum_{b\in {\rm n.n.}(a)}\!\! I_{ab} = \dot{\rho}_a$, where $\rho_a$  is the density operator  on site $a$. 
Therefore the gauge invariance condition for expectation values of the  magnetization density, Eq.~(\ref{eq:gaugeinvariance}), is satisfied 
if and only if the density on every site is stationary over the time-window from $0$ to $\tau$:  $\langle \dot{\rho}_a\rangle_\tau = 0$. 
This condition is the lattice-analogue of the condition that the current density in the continuum must be divergence-free.

\subsection{Ampere's law on the lattice}
To prove the lattice version of Ampere's law,  we first consider the case where 
the vector potential is given by $A_{ab}$  on a single bond $ab$, in the direction from site $b$  to site $a$, and zero everywhere else.
In this situation the magnetic flux is zero everywhere, except for the  two  plaquettes $p$  and $q$ adjacent to the bond $ab$, here taken such that the direction from site $b$  to site $a$  is  counterclockwise with respect to plaquette $p$.
In these two neighboring plaquettes, the fluxes are given by $\phi_p=A_{ab}$  and $\phi_q=-A_{ab}$, respectively.
Hence, with this choice of gauge (i.e., $A$ nonzero on a single bond), 
\be 
\frac{\partial H(t)}{\partial A_{ab}} =  \frac{\partial H(t)}{\partial \phi_p}- \frac{\partial H(t)}{\partial \phi_q}.
\label{eq:AmpereOpEq}
\ee
Noting that $\frac{\partial H(t)}{\partial A_{ab}} =-I_{ab}(t)$, and $m_p = -\frac{\partial H(t)}{\partial \phi_p}$, we obtain an operator equation similar to Eq.~(4)  in the main text. 
However, this {\it operator equation} holds only in the specific gauge above, where $A$ is nonzero only on the bond $ab$.
Importantly, as shown above, the {\it time-averaged expectation value} of the right hand side is gauge-independent 
for times $\tau$ where the density is stationary, $\langle \dot \rho\rangle_\tau= 0$.
Therefore Eq.~(\ref{eq:AmpereOpEq}) produces meaningful physical results, and reduces to Eq.~(4) of the main text, when it is used to compute time-averaged expectation values in stationary states.
\section{Relation to winding number}
Here we show that the quantized value of the magnetization density for a fully-localized Floquet system on a torus, $\bar m_\infty$, is a topological invariant; its value is equal to $W[U]/T$, where  $W[U]$ is the winding number introduced in Ref.~\onlinecite{AFAI}.
Noting that the numbers $W[U]$  and $\bar m_\infty$  do not change when we increase the system size, provided that all Floquet states remain localized, we  will consider  the limit where the size $L$ goes to infinity. 
In this section, we work in the Heisenberg picture.

In order to define the winding number $W[U]$, we consider the Hamiltonian   
$ H(\vec A, t)$  of the system  when a uniform vector potential $\vec A$ is introduced along the surface of the torus.
Let $U(\vec A,t)$  be the corresponding evolution operator of the system. 
As an important ingredient in the computation of the winding number, we first define the effective Hamiltonian of the system, $H_{{\rm eff},\, \varepsilon}(\vec A)$, 
   via: $U(\vec A,T)=e^{-iH_{{\rm eff}, \varepsilon}(\vec A)T}$, where the eigenvalues of $H_{{\rm eff},\varepsilon}(\vec A)$  lie in the interval $[\varepsilon,\varepsilon+2\pi/T)$. 
Here $\varepsilon$ is chosen within one of the system's quasienergy gaps, which are present due to the finite extent of the system for any fixed $L$ (see Ref.~\onlinecite{AFAI}). 
To find the system's winding number, we define the $2T$-periodic evolution $\tilde U_\varepsilon(\vec A,t)$,  obtained by first evolving the system with Hamiltonian $H(\vec A,t)$ in the time-interval $[0,T]$, and then applying a static Hamiltonian $-H_{{\rm eff}, \varepsilon}(\vec A)$  in the time-interval $[T,2T]$. 
The evolution operator $\tilde U_\varepsilon(\vec A,t)$  is given by  $U(\vec A,t)$  in the first half of the driving, from $0$  to $T$, and by $e^{-iH_{{\rm eff},\varepsilon}(\vec A)(2T-t)}$  in the second half of the driving.
In particular, the extended evolution satisfies $\tilde U_\varepsilon (\vec A,2T)=1$. 

With the definition of $\tilde U_\varepsilon(\vec A,t)$ above, we obtain the winding number of the evolution via: 
\begin{align}
W[ U] =
& \frac{1}{8\pi ^2 }  \int_0^{2T}\!  dt \, \int_{0}^{2\pi/L}\!\!\! d^2\! \vec{A}\, \notag \\ 
&\Tr \left(\tilde U^\dagger \dt \tilde U \cdot \tilde U^\dagger \partial _{A_x} \tilde U  \cdot    \tilde U^\dagger \partial_ {A_y}  \tilde U \right) 
- x \leftrightarrow y
\label{eq:WindingNumberDefinition}.
\end{align}
Given that $W$ is independent of $\varepsilon$ (see Ref.~\onlinecite{AFAI}), for brevity we drop the subscript $\varepsilon$ on $\tilde U$ here and below. 

As a first step in our derivation, we rewrite the above formula using basic identities for the time-evolution operator.
We first use the identities  $\dt \tilde U  =-i\tilde H \tilde U $ and $\partial _{A_x} \tilde U \cdot \tilde U^\dagger  = - \tilde U \partial _{A_x} \tilde U^\dagger$ to obtain 
\begin{align*}
W[ U] =
& \frac{i\epsilon_{\alpha\beta}}{8\pi ^2 }  \int_0^{2T}\!\!  dt \, \int_{0}^{2\pi/L}\!\!\! d^2\! \vec{A}\,\Tr \left( \tilde H  \partial _{A_\alpha} \tilde U  \cdot   \partial _{A_\beta} \tilde U^\dagger \right).
\end{align*}
Here $\epsilon_{\alpha\beta}$ is the antisymmetric tensor, with $\alpha,\beta = \{x,  y\}$.
Next, we perform partial integration over $A_\alpha$ and obtain 
\begin{align}
W[ U] =
& \frac{i\epsilon_{\alpha\beta}}{8\pi ^2 }  \int_0^{2T}\!  dt \, \left[
\int_{0}^{2\pi/L}\!\!\! \!d\! A_\beta
 \Tr\left(\tilde H  \tilde U  \cdot   \partial _{A_\beta} \tilde U^\dagger\right)^{A_\alpha=2\pi/L}_{A_\alpha=0} \right.
 \notag \\ 
&
 -
\left. \int_{0}^{2\pi/L}\!\!\! d^2\! \vec{A}\,
\Tr \left( \partial _{A_\alpha} \tilde H   \tilde U  \cdot   \partial _{A_\beta} \tilde U^\dagger\right)  \right].
\label{eq:WNPartialIntegration}
\end{align}
We now make use of the fact that we can write  $\tilde H(\vec A + \vec{\hat e}_\alpha 2\pi/L ,t)=X^\dagger_\alpha \tilde H(\vec A,t)X_\alpha $, where $\vec {\hat e}_\alpha$  is the $\alpha$-unit vector, and  $X_\alpha=e^{2\pi i x_\alpha/L}$   (see Ref.~\onlinecite{AFAI} for more details).
Similarly, $\tilde U(\vec A + \vec{\hat e}_\alpha 2\pi/L,t)= X^\dagger_\alpha\tilde U (\vec A,t)X_\alpha$. 
Using that $\partial _{A_{\beta}} X_{\alpha}=0$ when $\alpha\neq \beta$, together with the cyclic property of the trace, we obtain 
\begin{equation*}
\Tr \left(\tilde H \tilde U  \cdot   \partial _{A_\beta} \tilde U^\dagger\right)_{\vec A= \left(\frac{2\pi}{L},A_{\beta}\right)} =\Tr  \left(\tilde H \tilde U  \cdot   \partial _{A_\beta} \tilde U^\dagger\right)_{\vec A= \left(0,A_{\beta}\right)}.
\end{equation*}
Hence the integrand in the first term in Eq.~(\ref{eq:WNPartialIntegration})  vanishes, and 
\begin{align}
W[U] =
& \frac{-i\epsilon_{\alpha\beta}}{8\pi ^2 }  \int_0^{2T}\!  dt \, \int_{0}^{2\pi/L}\!\!\! d^2\! \vec{A} \Tr \left(\partial _{A_\alpha} \tilde H  \cdot  \tilde U\partial_ {A_\beta}\tilde U^\dagger\right) .
\end{align}
Using the identity $\partial _{A_\beta } \tilde U^\dagger= -\tilde U^\dagger \partial _{A_\beta } \tilde U \tilde  U^\dagger$,  along with  the cyclic property of the trace, we get
\begin{align*}
W[U] =
& \frac{i}{8\pi ^2 }  \int_0^{2T}\!  dt \, \int_{0}^{2\pi/L}\!\!\! d^2\! \vec{A} \Tr \left(\tilde U^\dagger \partial _{A_\alpha} \tilde H \tilde U   \cdot  \tilde U ^\dagger \partial_ {A_\beta }\tilde U\right) .
\end{align*}
Going to the thermodynamic limit $L\rightarrow \infty$,  we treat the integrand as constant within the $\vec A$-interval $[0,2\pi/L]$ (cf.~Ref.~\onlinecite{Hastings2015}).
Thus we arrive at the formula  
\begin{align}
W[U] =
& \frac{i}{2L^2 }  \int_0^{2T}\!  dt \, \Tr \left(\tilde U^\dagger_{} \left(\partial _{A_\alpha} \tilde H_{}\right)  \tilde U_{} \cdot   \tilde U^\dagger \partial_ {A_\beta}  \tilde U_{} \right) .
\label{eq:WindingNumberDefinitionTransformed}
\end{align}
What we have achieved so far, with Eq.~(\ref{eq:WindingNumberDefinitionTransformed}), is to relate the winding number to two Heisenberg picture operators, $\tilde U^\dagger \partial_ {\vec A}  \tilde U_{}$, and $ \tilde U^\dagger_{} \left(\partial _{\vec A} \tilde H_{}\right)  \tilde U_{} $.
Below we expose the physical meaning of each of these operators, and thereby link the winding number to the system's magnetization. 
%

\subsection{Displacement operator}
Having transformed the original winding number formula (\ref{eq:WindingNumberDefinition}) into the form of Eq.~\eqref{eq:WindingNumberDefinitionTransformed}, we now introduce an additional operator that will be useful in making the final connection with the magnetization.
Specifically, for a system with Hamiltonian $H(t)$, and evolution $ U(t)$, we introduce the ``displacement operator'' $\Delta {\vec{r}}(t)$: 
 \begin{eqnarray}
 \Delta {\vec{r}}(t)&\equiv &    -i U^\dagger (t) \partial _{\vec A}    U(t) .   
 \label{eq:DeltaRInitialDefinition}
\end{eqnarray}
With this definition, we note that $
 \dt \Delta {\vec{r}}(t)  =     U^\dagger(t) \left( - \partial _{\vec A }  H (t)\right)   U(t) $. 
The displacement operator  can be seen as the Heisenberg picture operator that measures  the 
displacement of a particle relative to its starting point, in the sense that displacement is the time-integral of the velocity.
This definition is important because the standard position operator on the torus is complicated by the necessity of imposing a branch cut due to the periodic boundary conditions.
The displacement operator in Eq.~(\ref{eq:DeltaRInitialDefinition}) is insensitive to this issue. 

To further elucidate the physical meaning of the displacement operator $\Delta \vec{r}(t)$, we consider the  case where the system has {\it open boundary conditions}, where the position operator $\vec{r}$ is naturally single-valued. 
In the lattice site basis, the Hamiltonian's matrix elements depend on the vector potential $\vec{A}$ in the following way:
\be 
H_{ab}(\vec A)=H_{ab} e^{i\vec A \cdot (\vec r_a -\vec r_b)}.
\ee
Consequently,  
$ \frac{\partial H(t)}{\partial \vec A} = i[\vec r,H(t)]$, 
and we find   $\dt \Delta \vec{r}(t) = \dt \vec r (t)$, where  $\vec r(t) = U^\dagger(t) \vec r U(t)$ is the time-evolved position operator in the Heisenberg picture. 
Using the initial condition $\Delta \vec r(0)=0$, we find 
\be 
\Delta \vec r(t) = \vec r(t) - \vec r(0).
\label{eq:delta_r_open}
\ee
For a system with periodic boundary conditions (e.g., a torus), it is not possible to write $\Delta \vec r(t)$ as a difference of initial and final positions, as in the above equation. 
However, when $\Delta \vec{r}(t)$  acts on a state $|\psi\rangle$  that stays localized within a region $S$ that is much smaller than the size of the torus, we can ignore the boundary conditions and write
\be 
\Delta \vec r(t) |\psi\rangle= (\vec r_S(t) - \vec r_S)|\psi\rangle,
\label{eq:DispOperatorPBC}
\ee
where $\vec{r}_S$  is a position operator defined with a branch cut outside $S$. 
(We note that the right-hand side does not depend on the exact location of the branch cut, as long as it is located far outside the region $S$.)

\subsection{Relationship with magnetization density}
Having defined the displacement operator, we now rewrite the winding number formula~(\ref{eq:WindingNumberDefinitionTransformed}) in terms of this operator. 
Using the definition in Eq.~(\ref{eq:DeltaRInitialDefinition}), we replace $\tilde U^\dagger \partial_{\vec A}\tilde U$  with $i \Delta \tilde{\vec{r}}(t)$, where $\Delta \tilde{\vec{r}}(t)$  is the displacement operator for the system governed by $\tilde H(t)$.
Similarly, as noted in the text below Eq.~(\ref{eq:DeltaRInitialDefinition}), we may replace $\tilde U^\dagger (\partial_{\vec A} \tilde H) \tilde U$ with $-\partial_t \Delta \tilde{\vec{r}}(t)$.
Thus we obtain
\be 
W[U] 
=\frac{1}{2L^2}\int_0^{2T} dt\,\Tr \left(\Delta \tilde{\vec{r}}(t) \times \partial _t \Delta\tilde{\vec{r}}(t)  \right) .
\label{eq:dispMag}
\ee
The integrand in Eq.~(\ref{eq:dispMag}) above has a very similar form to that of the magnetization, Eq.~(1) of the main text.
It remains to show that this expression, which involves the displacement operator defined in Eq.~(\ref{eq:DeltaRInitialDefinition}), precisely reduces to the magnetization discussed in the main text.

Writing out the trace in terms of the (localized) Floquet eigenstates $\{|\psi_n\rangle\}$, and using Eq.~\eqref{eq:DispOperatorPBC}, we obtain
\begin{align*}
W[U] 
&=\frac{1}{2L^2}\int_0^{2T} dt \sum_n \langle \psi_n| (\tilde{\vec r}_n(t)-\vec r_n) \times \partial _t \tilde{\vec r}_n(t) |\psi_n\rangle.
\end{align*}
Here  $\tilde{\vec r}_n(t)\equiv \tilde U^\dagger (t)\vec r_n \tilde U(t)$, where $\vec r_n$    is a position operator, defined with a branch cut far away from the region where the state $|\psi_n\rangle$  is localized. 
Using that $\tilde U(2T)=1$, such that $\tilde{\vec r}_n(2T)=\tilde{\vec r}_n(0) =\vec r_n$, we find
\begin{align}
W[U]=\frac{1}{2L^2}\int_0^{2T} dt \sum_n \langle \psi_n| \tilde{\vec r}_n(t)\times \partial _t \tilde{\vec r}_n(t) |\psi_n\rangle.
\label{eq:Wrcrossrdot}
\end{align}

In the first half of the driving, i.e., for $0 \le t \le T$, the system evolves according to the original Hamiltonian $H(t)$.
Here $\tilde{\vec r}_n(t)=\vec r_n(t)\equiv U^\dagger(t)\vec r_n U(t)$, where $U(t)$  is the corresponding evolution operator of the original system. 
In the second half of the driving, from $T$  to $2T$, the Hamiltonian of the system is given by $\tilde H_{}(t)=-H_{\rm eff}$, 
 and the time-evolution operator is given by $\tilde U(t)=e^{-iH_{\rm eff} (2T-t)}$. 
Using $\tilde{\vec r}_n(t) = \tilde U^\dagger(t) \vec r_n \tilde U(t)$, we then have (for $T \le t \le 2T$):
\begin{align*}
 \tilde{\vec r}_n(t)\times \partial _t \tilde{\vec r}_n(t) & =-i e^{iH_{\rm eff}(2T-t)} \vec r_n \times [\vec r_n, H_{\rm eff}]e^{-i H_{\rm eff}(2T-t)}. 
\end{align*}
Using $H_{\rm eff}=\sum_n P_n \varepsilon_n$, where $P_n = |\psi_n\rangle\langle \psi_n|$, we obtain  
\begin{align*}
 \langle \psi_n| \tilde{\vec r}_n(t)\times \partial _t \tilde{\vec r}_n(t)|\psi_n\rangle & =-i \sum_{m}\Bra{ \psi_n} \vec r_n \times [\vec r_n, P_m]|\psi_n\rangle \varepsilon_m. 
\end{align*}
Thus the integrand in Eq.~(\ref{eq:Wrcrossrdot}) is actually constant over the interval $T \le t \le 2T$.
This allows us to perform part of the integration and obtain
\begin{align}
W[U]&=\frac{1}{2L^2}\int_0^T dt\sum_n \langle \psi_n|\vec r_n (t) \times\dt \vec r_n(t)|\psi_n\rangle \notag \\
&+\frac{iT}{2L^2}\sum_{m,n}\langle \psi_n|\vec r_n\times [\vec r_n,P_m]|\psi_n\rangle\varepsilon_m.
\label{eq:Wtwoterms}
\end{align}

We now argue that the last term in Eq.~(\ref{eq:Wtwoterms}) must be zero. 
To do this,  we note that for a fully-localized system, the winding number is independent of the choice of the quasienergy zone (i.e., the position of the branch cut $\varepsilon$ in $H_{{\rm eff}, \varepsilon}$, see Ref.~\onlinecite{AFAI}).
If we shift the quasienergy cut to the gap between $\varepsilon_{m_0}$  and $\varepsilon_{m_1}$, where $\varepsilon_{m_0}$  and $\varepsilon_{m_1}$ are the lowest- and second lowest quasienergies, respectively,  the quasienergy $\varepsilon_{m_0}$ changes by $2\pi/T$, while all other quasienergies remain the same: $\varepsilon_{m_0}\rightarrow \varepsilon_{m_0} + 2\pi /T$.
%
The invariance of the left-hand side of  Eq.~(\ref{eq:Wtwoterms})   under this shift of  quasienergy zone implies that 
\be
\sum_n \langle \psi_n|\vec r_n\times [\vec r_n, P_{m_0}]|\psi_n\rangle=0.
\ee
Since the branch cut could be placed anywhere in the spectrum, 
the argument above should in fact hold for any  choice of $m_0$.
Therefore the last term in Eq.~(\ref{eq:Wtwoterms}) must vanish, and we arrive at 
\be 
W[U]=\frac{1}{2L^2}\int_0^T dt\sum_n \langle \psi_n|\vec r_n (t) \times\dt \vec r_n(t)|\psi_n\rangle. 
\ee
Following the discussion in the main text, we identify 
\be
\frac{1}{2 T}\int_0^T dt\, \langle \psi_n|\vec r_n (t) \times\dt \vec r_n(t)|\psi_n\rangle =  \langle M\rangle^{\!(n)}_T
\ee
as the time-averaged magnetization of Floquet eigenstate $n$. 
Hence 
\be 
W[U] = \frac{T}{L^2} \langle M \rangle_T,\quad \langle M \rangle_T=\sum_n \langle M\rangle^{\!(n)}_T,
\ee
where $\langle M \rangle_T$  is the total magnetization of the system when all states are occupied (on a torus). 
Using $\langle M \rangle _T = L^2 \bar m_{\infty}$, we finally  arrive at
\be 
\bar m_{\infty} = \frac{W[U]}{T}.
\ee
This is what we set out to show: the magnetization density of a fully-localized Floquet system is a topological invariant, with its value equal to the winding number identified in Ref.~\onlinecite{AFAI}, divided by the driving period, $T$.

\section{Measurement of magnetization in a cold atoms experiment}
\label{sec:Meausrement}
In this section, we prove Eq.~(10) in the main text. 
We show that the time-averaged magnetization can be measured via the net $y$-component of total (pseudo)-spin of a cloud of two-component cold atoms subjected to a spin-dependent artificial magnetic field. 
In this section, we will work in the Heisenberg picture.
For an individual atom in the experiment, the wave function before the measurement is given by 
\be 
|\psi\rangle =\frac{1}{\sqrt 2}|\chi\rangle\otimes\left(|\! \uparrow\rangle +|\!\downarrow\rangle\right),
\ee
where $|\chi\rangle$ denotes the orbital part of the atom's wave function, and the tensor product separates the orbital and spin parts of the wave function.
The time evolution operator of the system for the case where the spin-dependent effective field acts only on the $\Ket{\! \uparrow}$ spin component is given by 
\be
\mathbb U(\tau) = U_B(\tau) \otimes |\!\!\uparrow\rangle\langle\uparrow\!\!|+U_0(\tau) \otimes|\!\!\downarrow\rangle\langle\!\! \downarrow\! \!|,
\ee
 where $U_B(\tau)$  is the 
 time-evolution operator (acting only on the system's orbital degrees of freedom) when a uniform field $B$  is applied.  

After an evolution time $\tau$ in the presence of the effective field $B$, the atom's wave function is given by 
\be 
|\psi (\tau)\rangle =\frac{1}{\sqrt 2}\left(U_B(\tau)|\chi\rangle\otimes|\! \uparrow\rangle +U_0(\tau)|\chi\rangle\otimes |\!\downarrow\rangle\right).
\ee
Hence, at time $\tau$, the expectation value of the $y$-spin operator 
$ 
\sigma_y=\frac{i}{2}(|\!\!\uparrow\rangle\langle \downarrow \!|-|\!\!\downarrow \rangle\langle \uparrow\!\!|)
$
is given by 
\be 
\langle \sigma_y(\tau)\rangle =\frac{i}{2} \langle \chi|\left(U^\dagger_B(\tau) U_0(\tau)- U^\dagger_0(\tau) U_B(\tau)\right)|\chi\rangle.
\label{eq:SpinYExpValue}
\ee 
Using $U_B(\tau) =U_0(\tau)+ B\frac{\partial }{\partial B} U_B(\tau)\vert_{B = 0}+ \mathcal O(B^2)$, valid in the linear response regime of weak fields, we obtain 
\begin{align}
  \langle \sigma_y(\tau)\rangle =   -i B \langle \chi|\left(U^\dagger_0(\tau) \frac{\partial }{\partial B} U_0(\tau)\right)|\chi\rangle  + \mathcal O (B^2),
%
\label{eq:UdBU}
\end{align}
where for brevity we write $\frac{\partial}{\partial B} U_B(\tau)\vert_{B = 0} \equiv \frac{\partial}{\partial B} U_0(\tau)$. 
To arrive at Eq.~(\ref{eq:UdBU}), we used the identity $ \frac{\partial }{\partial B} U^\dagger_{0}\cdot U_0 =- U^\dagger_{0}\cdot \frac{\partial }{\partial B} U_0$.
Using Eq.~\eqref{eq:QEDeriv2} we obtain the following result, which is valid on short times where the spin precession angle remains small: 
\be 
\langle \sigma_y(\tau)\rangle = 
  B\int_0^\tau dt\, \langle \chi (t)| M(t)|\chi (t)\rangle 
 + \mathcal O(B^2).
\label{eq:UdBU_int}
\ee
Here we have introduced the operator $M(t)$ as a shorthand for $-\frac{\partial H(t)}{\partial B}$.
We note that this operator, and its expectation values (for non-stationary states), in general depend on the implementation of the gauge field, see discussion below.

The above result, Eq.~(\ref{eq:UdBU_int}), holds for an individual atom. 
For a droplet of many non-interacting atoms 
the droplet's total $y$-spin $\langle S_y\rangle$ can be obtained by summing together their individual contributions:
\be 
\langle S_y(NT)\rangle =  B  NT  \sum_{j} \langle M\rangle^{(j)}_{NT} + \mathcal O (B^2),
\label{eq:TotalMagnetizationYSpinRelation}
\ee
where the sum runs over all atoms $j$  in the droplet, 
 and $\langle M\rangle_\tau^{(j)}$ denotes the time-averaged expectation value of $M(t)$ for the atom $j$, taken over the interval $0 \le t \le \tau$. 
Importantly, for long times, $N\rightarrow\infty$, the particle density is stationary and $\langle M\rangle_{NT}^{(j)}$ becomes gauge independent.
In this limit, $ \sum_{j} \langle M\rangle^{(j)}_{NT}\rightarrow \doubleAvg{M}$ and we find 
\be
\label{eq:asymptoticValue}\lim_{NT\rightarrow\infty}\frac{1}{BNT}\langle S_y(NT)\rangle =  \doubleAvg{M} + \mathcal O (B).
\ee

For a finite number of periods $N$, there will in general be a transient correction to the relation in Eq.~(\ref{eq:asymptoticValue}) above. 
Consider a filled droplet, as described in the main text, where the many-body state is described by a single Slater determinant.
Within such a state, atoms localized deep inside the bulk of the droplet (i.e., centered many localization lengths from its boundary), where all sites are filled, can be taken to be occupying Floquet eigenstates.
For an atom $j$ initialized in a Floquet eigenstate $n$, $\langle M\rangle_{NT}^{(j)} = -\frac{\partial \varepsilon_n}{\partial B}$
 for any integer number of periods, $N$. 
Thus atoms in the bulk do not give any transient corrections to Eq.~(\ref{eq:asymptoticValue}). 
However, an atom $j$ localized near the boundary of the droplet does not generically occupy a single Floquet eigenstate. 
In this case, the contribution of atom $j$ to the total density is not stationary over a single period, and $\langle M\rangle^{(j)}_{NT}$   generally depends on $N$. 
Thus the motion of atoms localized in a strip of width $\sim \xi$ along the boundary of the droplet produces a transient deviation of $\frac{1}{BNT}\avg{S_y(NT)}$ from its long-time asymptotic value $\doubleAvg{M}$.

The non-universal transient depends on details of the implementation, including in particular the choice of ``gauge'' used for producing the effective spin-dependent magnetic field.
That is, the spin rotation of an atom moving through the lattice depends explicitly on the ``vector potentials'' $\vec{A}_\uparrow$ and $\vec{A}_\downarrow$ for up and down spins, respectively, and not only on the effective magnetic fields $\vec{B}_\uparrow = \nabla \times \vec{A}_\uparrow$ and $\vec{B}_\downarrow = \nabla \times \vec{A}_\downarrow$.
Independent ``gauge'' transformations of $\vec{A}_\uparrow$ and $\vec{A}_\downarrow$ correspond to position-dependent spin rotations around the $z$-axis.
Since the system is initialized and measured in a fixed, spatially uniform frame, there is no symmetry under spin-dependent gauge transformations.
We now estimate the magnitude of the transient correction.
To do so, we consider the case of a circular droplet of radius $R$, where the magnetic field is implemented in the symmetric gauge (here the origin of the coordinate system is located in the droplet's center).
In the symmetric gauge, recall from Sec.~\ref{sec:MagnetizationHamiltonianRelationship} that $
M(t) = -\frac{\partial H(t)}{\partial B} = \frac{1}{2} \vec {\hat z} \cdot (\vec r\times \dot{\vec r}(t))$. 
For an atom at the boundary of the droplet we write $\vec{r}(t) = \vec{R} + \delta\vec{r}(t)$, where $\vec{R} = \doubleAvg{\vec{r}(t)}$ is a vector of length $\sim R$ pointing from the origin to the atom's long-time-averaged position, and $\delta\vec{r}(t)$ describes the motion around this point, with $|\delta\vec{r}| \sim \xi$.
The time-averaged expectation value of $M$  
 for an atom in the boundary region is then
 \be 
 \langle M\rangle^{\! (j)} _{NT}= \frac{1}{2} \vec{\hat z }\cdot \left[\vec R \times \langle \delta \dot{\vec{r}}\rangle_{NT} + \langle  \delta \vec r \times  \delta \dot{\vec{r}}\rangle_{NT}\right].
 \label{eq:MagnetizationDecomposition}
 \ee
The first term yields a contribution to $\avg{M}_{NT}$ of order $R \avg{\dot{r}_\parallel}_{NT}$, where
$\dot{r}_{\parallel}(t)$ denotes the tangential component of the atom's velocity along the boundary. 
Since the atom must remain confined within a region of linear dimension $\xi$ for all times, the $N$-period average of the tangential velocity takes a typical value of order $\xi/NT$.
Therefore we expect the corresponding transient contribution to $\langle M\rangle_{NT}^{(j)}$ to have a magnitude at most $\sim \frac{R\xi}{N T}$. 
Assuming that the atoms are initially randomly distributed within their respective localization areas (this is assured by letting particle density in the droplet reach a steady profile 
 before the measurement begins), the {\it sign} of $\avg{\dot{r_\parallel}}_{NT}$ is expected to be random. 
Any transient contributions to $\langle M\rangle^{\!(j)}_{NT}$  from the second term in Eq.~\eqref{eq:MagnetizationDecomposition} involving $\delta\vec{r}\times \delta \dot{\vec{r}}$ are expected to be relatively suppressed by a factor $\xi/R$, and we ignore them below.

Having estimated the scale of the transient contribution to $\avg{M}^{(j)}_{NT}$ for each boundary atom, we now infer the net contribution of all atoms to the net transient deviation of $\frac{1}{BNT}\avg{S_y(NT)}$ from the asymptotic value $\doubleAvg{M}$.
First, note that total number of atoms in the boundary region (a strip of width $\xi$ around the perimeter of the droplet) is of order  $R \xi /a^2$. %
Assuming a random sign for the contribution of each atom, we get a net transient correction with magnitude of order $\sqrt{\frac{R\xi}{a^2}}\cdot \frac{R \xi }{ NT}$. 
Using $A_{\rm loc}= \xi^2$, and $A_{\rm filled} \sim R^2$, we thus obtain
\be 
\sum_j \langle M\rangle^{(j)} _{NT} =  \doubleAvg{M} + \frac{1}{NT} \mathcal O \left(\frac{A_{\rm loc} A_{\rm filled}}{a \sqrt{R \xi}} \right).
\label{eq:CloudMagnetizationResult}
\ee
While this result was obtained for a 
field implemented in the symmetric gauge, analogous arguments to those above can be used  for other natural implementations, e.g. the Landau gauge,  to show that the transient should have the same magnitude as above. 

Using Eq.~\eqref{eq:CloudMagnetizationResult} in Eq.~\eqref{eq:TotalMagnetizationYSpinRelation}, we see that
\be 
\label{eq:SyMag} 
\frac{{\langle  S_y(NT) \rangle }}{BNT}  =\doubleAvg{M}    + \frac{1}{NT} \mathcal O \left(\frac{A_{\rm loc} A_{\rm filled}}{a \sqrt{R \xi}}\right)+ \mathcal O(B).
\ee
Hence the  cloud's total magnetization can be extracted from  
the asymptotic behaviour of the growth rate of $\langle S_y(\tau)\rangle$ in the long-time limit.
The result for the average $y$-spin {\it per particle} $\avg{\overline{\sigma}_y(NT)}$, in Eq.(10) in the main text, is obtained by dividing both sides of Eq.~(\ref{eq:SyMag}) with the total number of atoms, $A_{\rm filled}/a^2$.

The ``long time limit'' in which the magnetization can be extracted should be understood as a time that is long compared with the damping of transients due to the system's initialization, but still short enough that the
 atoms' spin precession angle is small. 
The necessary separation of timescales can be guaranteed both by working at small fields, $B$, and by taking a large enough droplet (since the transient correction to $\avg{\overline{\sigma}_y(NT)}$ decays as $1/\sqrt{R}$).
  In practice, our numerics show that the transients can be made quite small for square droplets of only a few tens of lattice sites per side (see below and main text).

\section{Numerical simulation}
\begin{figure}[t!!]
\includegraphics[scale=0.6]{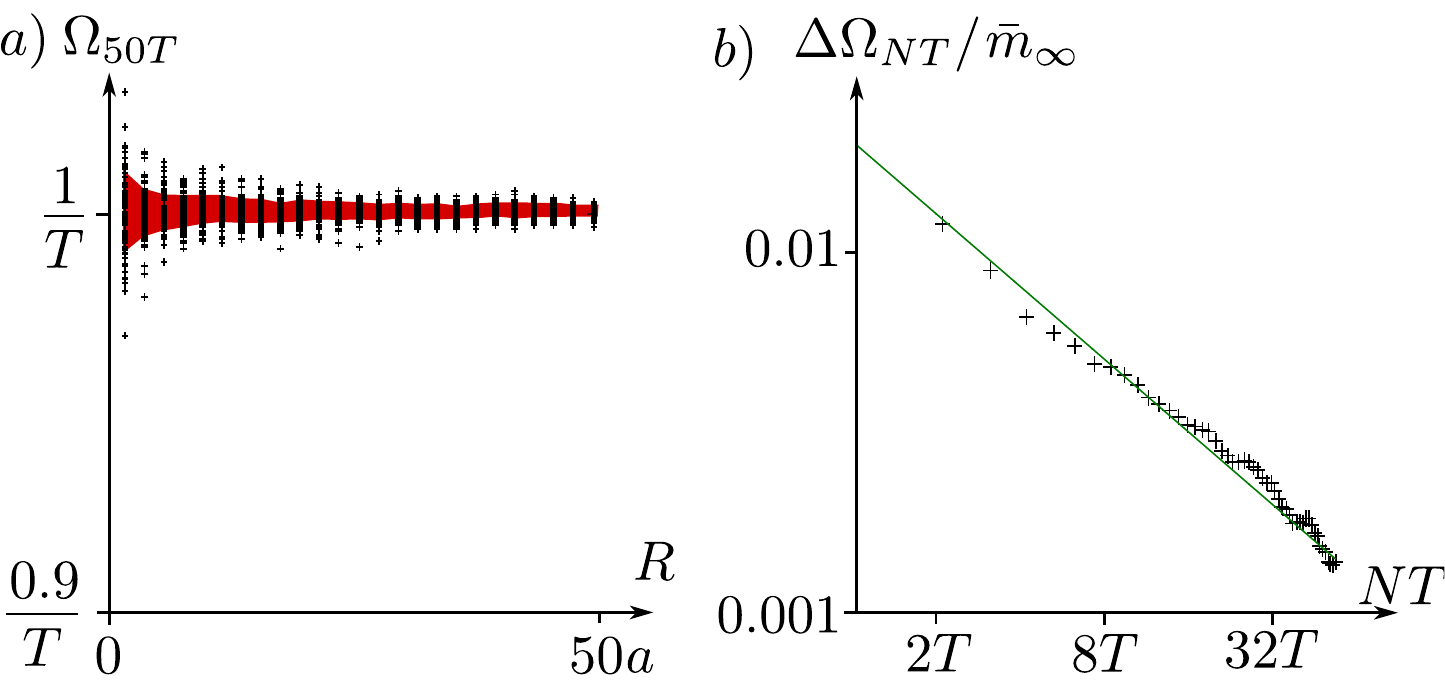}
\caption{Statistical behaviour of the normalized growth rate $\Omega_{NT}$
, whose saturation value yields the long-time-averaged magnetization density. 
a)  Normalized growth rate $\Omega_{NT}$  as function of droplet size $R$,  
obtained for 100 disorder realizations, with parameters set as in the main text (for each $R$, each realization corresponds to one black cross).
The red shading indicates the
interval within one standard deviation from the data points' mean. 
b) 
Deviation $\Delta \Omega_{NT}$  of the net $y$-spin growth rate from the expected saturation value $\bar{m}_\infty = 1/T$, as a function of the averaging time $NT$, 
 taken as an rms-average over 100 disorder realizations. 
The data are shown in a logarithmic plot. 
}
\label{fig:Statistics}
\end{figure}

Here we provide additional details from the numerical simulations, beyond what was discussed in the main text.
The magnetic field in the simulation was implemented in the Landau gauge, $\vec A=(0,-B(x-x_0))$, where $x_0$  is located in the center of the lattice.

To explore the generic behavior of the system in 
the parameter regime used in the main text, 
  we find and diagonalize the Floquet operator for 100 random 
  disorder realizations, on a lattice of $80\times 80$  sites with periodic boundary conditions.
 Among all Floquet eigenstates across these 100 realizations, we find the largest localization length 
 to  be $11.7 a$, where $a$  is the lattice constant.
 Thus we are well within the fully-localized, AFAI regime.
 We furthermore have compiled statistics to demonstrate how the normalized growth rate $\Omega_{NT}\equiv \frac{1}{Ba^2 NT}\langle \overline{\sigma}_y(NT)\rangle$  converges to the quantized value with system size and averaging time, which we now discuss.

In Fig.~\ref{fig:Statistics}a we show the time-averaged magnetization density after $50$  periods as function of $R$ (the side length of the filled squared droplet) for each of the $100$ realizations.
For each value of $R$, each black cross indicates the the value obtained for a specific realization.
The red area marks the interval within one standard deviation from the mean value of $\Omega_{50 T}$, obtained from the $100$  realizations. 
For \textit{all} disorder realizations we see that $\Omega_{50T}$ rapidly converges to the quantized value as the size of the filled region, $L$, is increased.

To see how  the average magnetization converges to the quantized value with the averaging time, $NT$, we investigate the deviation $\Delta \Omega_{NT}$ of $\Omega_{NT}$   from the quantized value $\bar{m}_\infty = 1/T$ as a function of $N$.
The value of $\Delta \Omega_{NT}$ is obtained as a root-mean-squared deviation, taken over the $100$ realizations, in the case where a region of $50\times 50$  sites is initially occupied.
The data are shown in a log-log plot in Fig.~\ref{fig:Statistics}b.
The linear trend indicates that the deviation decreases with a power-law scaling behaviour.
From  a linear fit (green line), we find  that the deviation  from the quantized value decreases  as $(NT)^{-0.64}$.



\begin{thebibliography}{53}
\bibitem[{foo({\natexlab{a}})}]{footnote:origin_invariance}
\bibinfo{note}{The orbital magnetization, Eq. (1), is independent of shifts of
  origin $\vec{r} \rightarrow (\vec{r}-\vec{r}_0)$ when evaluated in stationary
  states with \addFN{$\avg{\dot{\vec{r}}}_\tau = 0$}. In a fully localized
  system, this implies that the magnetization of a Floquet eigenstate averaged
  over an integer number of driving periods is origin-independent.}
\bibitem[{FN:({\natexlab{b}})}]{FN:MagneticFieldUnits}
\bibinfo{note}{We measure magnetic field in units of [1/Area], such that the
  Aharonov-Bohm phase of a closed trajectory is equal to the flux enclosed by
  the path.}
\bibitem[{foo({\natexlab{b}})}]{footnote-stationary}
\bibinfo{note}{Although the particle density is not strictly stationary,
  localization implies $\lim_{\tau\to\infty}\left< \dot{\rho}\right>_{\tau}=0$.
  Therefore the long time average $\left< \!\left< m_p \right>\!\right>$ is
  gauge invariant~\cite{FN:MagnetizationDensityGaugeInvariance} and obeys
  Ampere's law, Eq.~(\ref{eq:AmperesLaw}).}
\bibitem[{foo({\natexlab{c}})}]{footnote:DiracQuantization}
\bibinfo{note}{This follows from the Dirac quantization condition: the torus
  can enclose an integer number of magnetic monopoles, each with a quantized
  charge of $\Phi_0$.}
\bibitem[{foo({\natexlab{d}})}]{footnote:labeling}
\bibinfo{note}{The labeling of Floquet eigenstates in the presence of the
  uniform field $B_0$ is defined such that $\Amp{\psi_n(t,B_0)}{\psi_m(t, 0)} =
  \delta_{nm} + \mathcal{O}(1/A)$. For large systems, this prescription holds
  for all but an exponentially small subset of disorder realizations.}
\bibitem[{foo({\natexlab{e}})}]{footnote:InitialCondition}
\bibinfo{note}{Here we choose $\log |U(t)|$ to be continuous in time, with
  $\log|U(0)| = 0$.}
\end{thebibliography}

\end{document}